\PassOptionsToPackage{unicode=true}{hyperref} 
\PassOptionsToPackage{hyphens}{url}
\documentclass[]{article}
\usepackage{lmodern}
\usepackage{amssymb,amsmath}
\usepackage{ifxetex,ifluatex}
\usepackage{fixltx2e} 
\ifnum 0\ifxetex 1\fi\ifluatex 1\fi=0 
  \usepackage[T1]{fontenc}
  \usepackage[utf8]{inputenc}
  \usepackage{textcomp} 
\else 
  \usepackage{unicode-math}
  \defaultfontfeatures{Ligatures=TeX,Scale=MatchLowercase}
\fi
\IfFileExists{upquote.sty}{\usepackage{upquote}}{}
\IfFileExists{microtype.sty}{%
\usepackage[]{microtype}
\UseMicrotypeSet[protrusion]{basicmath} 
}{}
\IfFileExists{parskip.sty}{%
\usepackage{parskip}
}{
\setlength{\parindent}{0pt}
\setlength{\parskip}{6pt plus 2pt minus 1pt}
}
\usepackage{hyperref}
\hypersetup{
            pdftitle={PAT: a new algorithm for all-gather and reduce-scatter operations at scale},
            pdfauthor={Sylvain Jeaugey, NVIDIA},
            pdfborder={0 0 0},
            breaklinks=true}
\usepackage{graphicx,grffile}
\makeatletter
\def\maxwidth{\ifdim\Gin@nat@width>\linewidth\linewidth\else\Gin@nat@width\fi}
\def\maxheight{\ifdim\Gin@nat@height>\textheight\textheight\else\Gin@nat@height\fi}
\makeatother
\setkeys{Gin}{width=\maxwidth,height=\maxheight,keepaspectratio}
\providecommand{\tightlist}{%
  \setlength{\itemsep}{0pt}\setlength{\parskip}{0pt}}
\ifx\paragraph\undefined\else
\let\oldparagraph\paragraph
\renewcommand{\paragraph}[1]{\oldparagraph{#1}\mbox{}}
\fi
\ifx\subparagraph\undefined\else
\let\oldsubparagraph\subparagraph
\renewcommand{\subparagraph}[1]{\oldsubparagraph{#1}\mbox{}}
\fi

\makeatletter
\def\fps@figure{htbp}
\makeatother

\usepackage{float}
\title{PAT: a new algorithm for all-gather and reduce-scatter operations at
scale}
\author{Sylvain Jeaugey, NVIDIA}
\date{}

\begin{document}
\maketitle
\begin{abstract}
This paper describes a new algorithm called PAT, for Parallel Aggregated
Trees, and which can be used to implement all-gather and reduce-scatter
operations. This algorithm works on any number of ranks, has a
logarithmic number of network transfers for small size operations,
minimizes long-distance communication, and requires a logarithmic amount
of internal buffers, independently from the total operation size. It is
aimed at improving the performance of the NCCL
{[}\protect\hyperlink{ref-nccl}{1}{]} library in cases where the ring
algorithm would be inefficient, as its linear latency would show poor
performance for small sizes and/or at scale.
\end{abstract}

\hypertarget{all-gather-and-reduce-scatter-algorithms}{%
\section{All-gather and reduce-scatter
algorithms}\label{all-gather-and-reduce-scatter-algorithms}}

The NCCL library currently only implements the \emph{Ring} algorithm for
all-gather and reduce-scatter operations. The \emph{Tree} algorithm has
provided logarithmic latency to all-reduce operations for a long time,
but that algorithm doesn't apply to all-gather and reduce-scatter
operations.

Logarithmic algorithms exist for all-gather in the literature, and
inside MPI implementations. In this paper, we'll mostly talk about the
\emph{Bruck} {[}\protect\hyperlink{ref-bruck}{2}{]} algorithm and the
\emph{Recursive doubling} algorithm
{[}\protect\hyperlink{ref-thakur}{3}{]}. Both work perfectly in theory,
but in practice, they do not perform well, because their last steps
consist in every rank sending a lot of data to very distant ranks, often
crossing many levels of network switches. With these algorithms, at
every step, we double the distance we cross, and we also double the
amount of data we send, as we gather more and more data. The last step
sees every rank send half of the total size to its most distant rank. On
large fabrics, that last step frequently runs many times slower than the
theory due to static routing, or due to higher levels of the fabric
being tapered.

Another downside of the \emph{Recursive doubling} algorithm is that it
only works on number of ranks which are powers of two. This is a
significant constraint which we deemed not acceptable given a large
portion of the AI use cases do not use a power of two as their
data-parallelism dimension.

The \emph{Bruck} algorithm was originally an all-to-all algorithm, but
given we can implement all-gather as an all-to-all operation, it quickly
became a classic all-gather algorithm.

Both the \emph{Bruck} and \emph{Recursive doubling} algorithms rely on
the user-provided receive buffer acting as intermediate buffer to store
data before sending it to other ranks. This only works for all-gather
where the receive buffer is larger and can be written to; MPI semantics
do not allow the library to overwrite the send buffer, which is why
those algorithms were never used to implement reduce-scatter on large
sizes, despite reduce-scatter being a mirror operation to all-gather.

\hypertarget{binomial-tree-based-algorithms}{%
\section{Binomial-tree based
algorithms}\label{binomial-tree-based-algorithms}}

To understand the PAT algorithm, it is important to first understand how
the \emph{Bruck} algorithm works.

The \emph{Bruck} algorithm can be seen as implementing a binomial tree
for each rank. Data is broadcast from each rank to all other ranks
following a binomial tree, shifted for each rank. On the first step,
each rank will initiate the binomial tree of their own data, but every
step after that will also execute part of the binomial tree of other
ranks, aggregating more and more data at each step.

Note that the \emph{Recursive doubling} algorithm follows a similar
principle, except binomial trees from each rank are mirrored on the
dimensions of a n-dimension cube (which is why we need a power-of-two
number of ranks) rather than shifted.

Here is one description of how the \emph{Bruck} algorithm for all-gather
operations works:

\begin{enumerate}
\def\labelenumi{\arabic{enumi}.}
\item
  Send my data to the next rank, i.e.~send data for rank myrank-0 to
  myrank+1. Receive data from myrank-1.
\item
  Send data from myrank-1 and mine to myrank+2, receive data for
  myrank-3 and myrank-2 from myrank-2.
\item
  Send data from myrank-3, myrank-2, myrank-1 and myrank-0 to myrank+4,
  receive data for myrank-7 to myrank-4 from myrank-4.
\end{enumerate}

And more generally, send data for ranks
myrank-(2\textsuperscript{n}-1)..myrank to myrank+2\textsuperscript{n}
and receive data for ranks
myrank-(2\textsuperscript{(n+1)}-1)..myrank-2\textsuperscript{n} from
myrank-2\textsuperscript{n}. In this paper, we will refer to n as the
dimension, which is the power of two we're communicating with.

\begin{figure}
\centering
\includegraphics{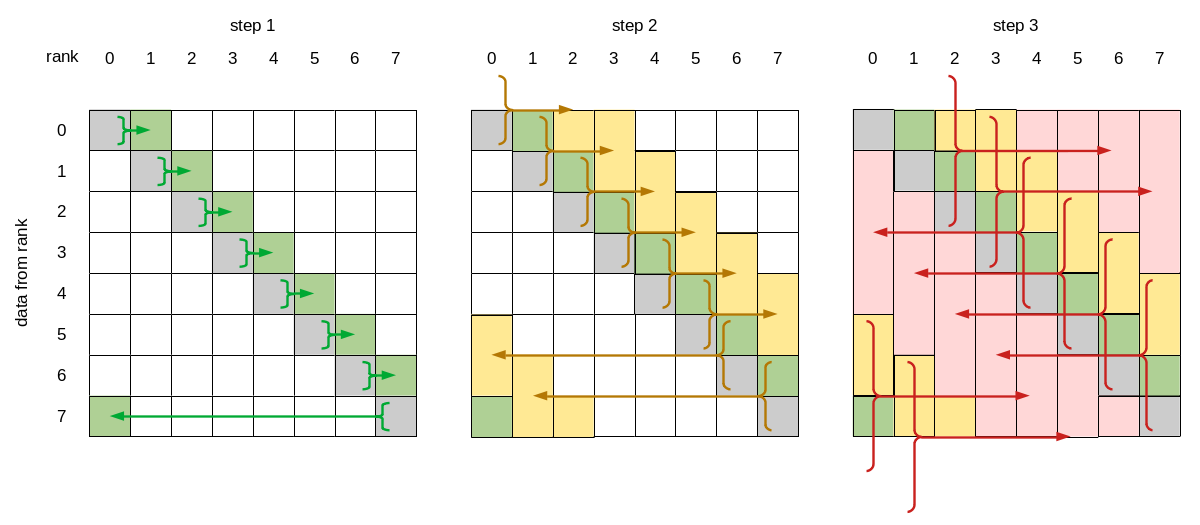}
\caption{Bruck algorithm}
\end{figure}

If we now consider separately the data from each rank, and look at how
that data gets broadcast, we can confirm that it follows a binomial
tree. In the global algorithm, steps from different binomial trees get
aggregated together to end up with a logarithmic total number of steps.

\begin{figure}
\centering
\includegraphics{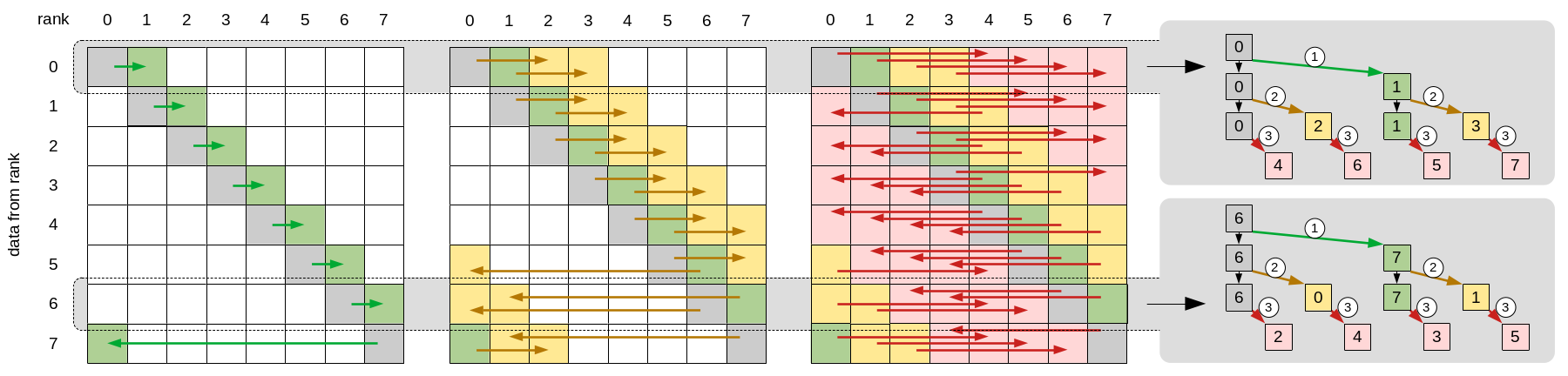}
\caption{Bruck algorithm, broadcasting each rank's data through a
binomial tree}
\end{figure}

To avoid sending a lot of data long-distance, we could reorganize the
dimensions to perform the farthest dimensions first and the nearest
dimensions last. However, the data we send to a given peer is no longer
contiguous and therefore require either some packing/unpacking, or to
send a linear number of messages, which can end up with a linear
latency, like rings.

\begin{figure}
\centering
\includegraphics{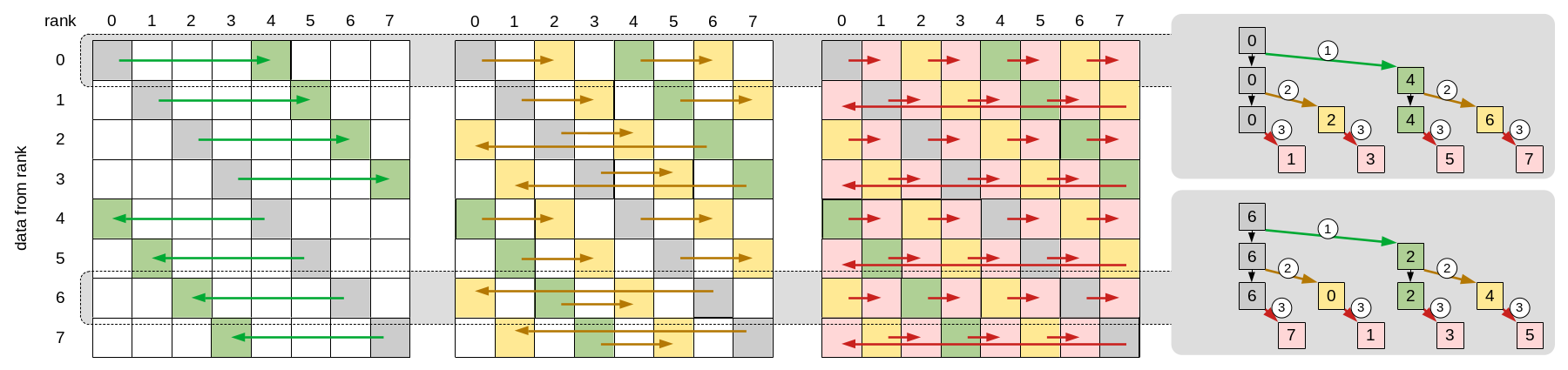}
\caption{Bruck algorithm, with farthest dimension first}
\end{figure}

Note that on the third step, each rank sends 4 chunks of data, coming
from 4 ranks, hence each ranks executes one step from 4 different
binomial trees. Conversely, the 4 send/recv operations we see in the
third step of any binomial tree are performed by 4 different ranks. In a
way, communication steps happen orthogonally to the binomial trees.

If the number of ranks is not a power of two, then we simply follow a
truncated binary tree. In the example illustrated below with 7 ranks,
the first step will send one chunk of data corresponding to its local
data, the second step will send 2 chunks (its local data, and the data
of rank-4), and the third step will send 3 chunks (its local data, the
data from rank-4 and the data from rank-6).

\begin{figure}
\centering
\includegraphics{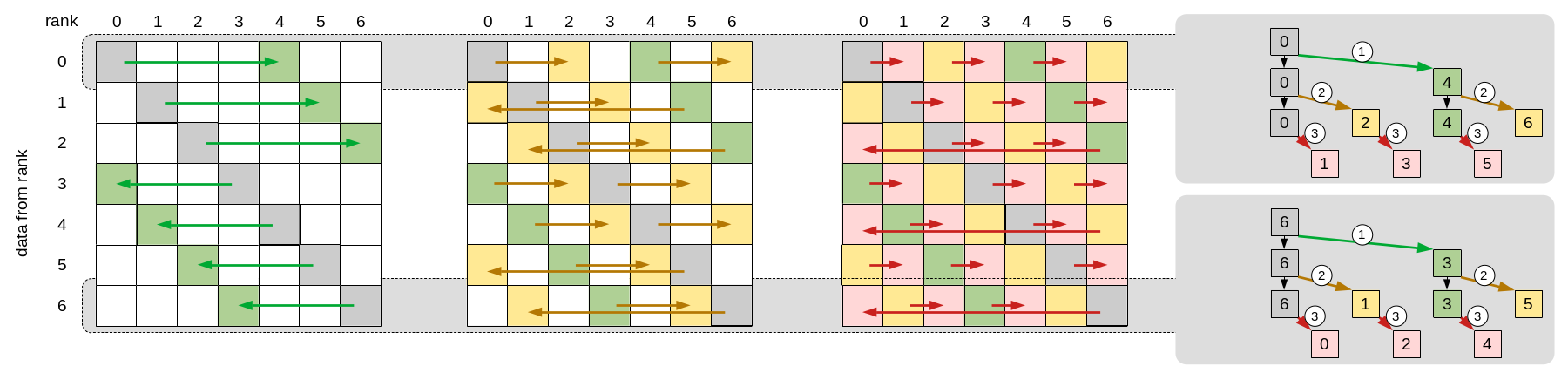}
\caption{Bruck algorithm, with nearest dimension first on a
non-power-of-two number of ranks}
\end{figure}

Again, the 3 chunks one rank sends are not the three steps shown in any
single binary tree. These are 3 chunks from 3 binary trees where the
rank is doing its part of that binary tree. For example, in the last
step, rank 0 will send to rank 1 its own data, data from 6, and data
from 4 (binomial tree for 4 is not shown, but we can see the 0 to 1
transfer in the binomial trees for 0 and 6). Orthogonally, on the last
step, ranks 0, 2 and 4 will send data respectively to 1, 3 and 5,
thereby completing the last step of the binomial tree for rank 0.

\hypertarget{the-pat-algorithm}{%
\section{The PAT algorithm}\label{the-pat-algorithm}}

The PAT algorithm was developped for both all-gather and reduce-scatter,
assuming that one operation being the mirror of the other, the same
algorithm would apply to both cases. This section will describe the
allgather version of PAT, as it can be derived from the Bruck algorithm,
and we will present the reduce-scatter version in a later section.

The PAT algorithm starts from the reversed-dimension Bruck algorithm
previously described, but additionally limits the amount of intermediate
buffering needed.

There are two main reasons why we may want to use intermediate buffers:

\begin{itemize}
\tightlist
\item
  We are performing a reduce-scatter operation and the receive buffer is
  limited in size.
\item
  We cannot directly send the user buffer to another rank (or receive
  data directly to the user buffer). Being able to do so may require to
  register that buffer onto the NIC or through other mechanisms to make
  the memory visible to the remote ranks, and can have a very high
  overhead.
\end{itemize}

In those cases, we may want to pre-allocate a dedicated buffer and map
it between ranks. The size of that buffer will be limited though.

The PAT algorithm will therefore adapt the amount of data it sends to
fit within the size of the intermediate buffer. If the maximum amount of
data we send to a given rank fits inside the intermediate buffer, it
will look exactly like the reversed-dimension Bruck algorithm.

However, in the case where we would not have enough space to hold the
maximum amount of aggregated data, we will limit the aggregation to the
maximum number of sub-trees our buffer size permits, and execute a
linear schedule within those sub-trees.

For example, in the case of 8 ranks where we can only aggregate the data
for two ranks, the algorithm would look like this:

\begin{figure}
\centering
\includegraphics{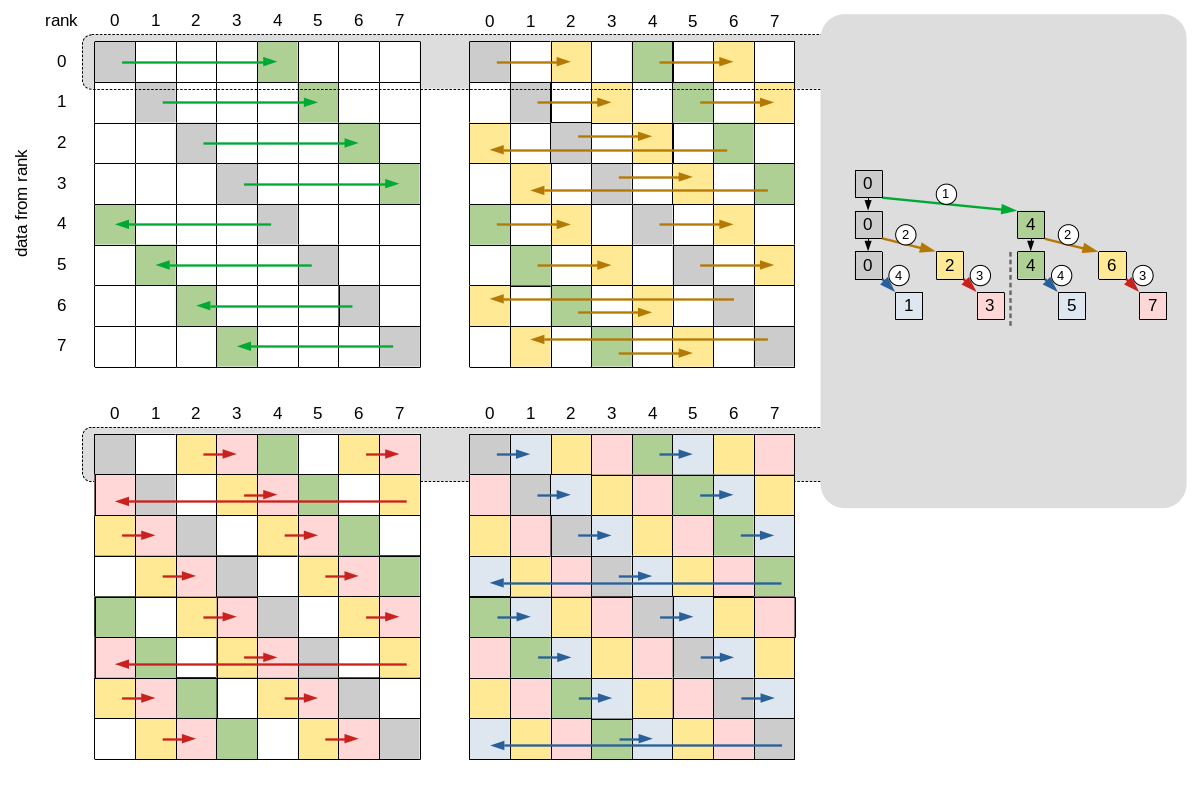}
\caption{PAT algorithm with 8 ranks, limiting aggregation to 2}
\end{figure}

The beginning of the algorithm is similar to Bruck, except we don't have
enough buffers for a single red step. We therefore split it into two
parts (red and blue). We perform the far step first to empty any
intermediate buffer we may need to reuse later, then perform steps which
are closer.

In the algorithm, we distinguish different phases: the first steps are
the top of the tree with full aggregation, but once we reach the maximum
aggregation level of two, we continue with only two parallel trees,
executing a linear number of steps to complete the execution of each
tree. In the example below, we have 1 step at the top (fully aggregated
logarithmic steps), and 3 steps within the tree (linear).

\begin{figure}
\centering
\includegraphics{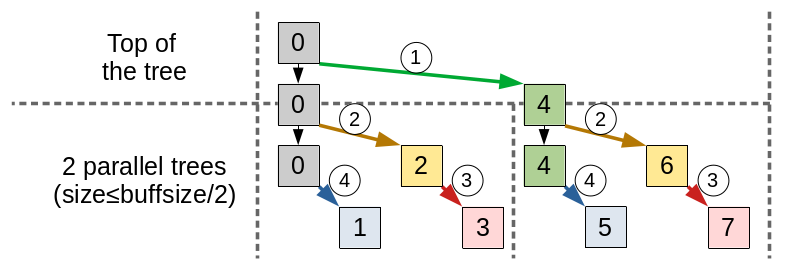}
\caption{PAT tree, limiting aggregation to 2}
\end{figure}

As the size of the operation increases, we will reduce the size of the
logarithmic part and increase the size of the linear part. This should
not be a problem for performance, given every transfer in the linear
part is performed with full buffers, which should guarantee close to
peak bandwidth.

The next 3 figures show how we transition from 8 trees to 2, in the case
of 16 ranks, and as the size per rank increases.

\begin{figure}
\centering
\includegraphics{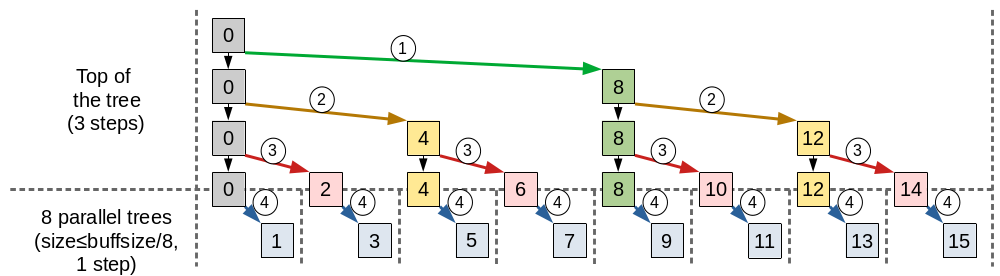}
\caption{PAT tree with 16 ranks and 8 trees, equivalent to
dimension-reversed Bruck}
\end{figure}

\begin{figure}
\centering
\includegraphics{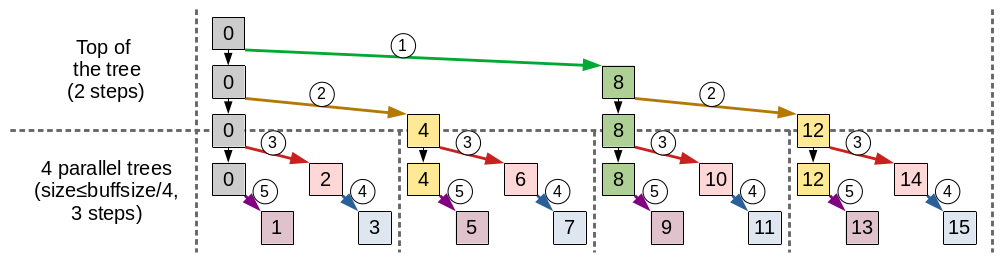}
\caption{PAT tree with 16 ranks and 4 trees}
\end{figure}

\begin{figure}
\centering
\includegraphics{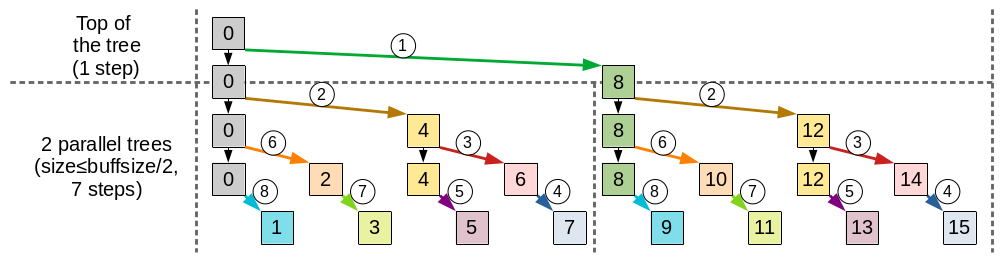}
\caption{PAT tree with 16 ranks and 2 trees}
\end{figure}

When the size per rank is larger than the buffer size, then we end up
with a single tree which has a linear number of steps. The algorithm
starts by sending data far, then progressively getting closer to the
root. This has a fundamental property which guarantees that we will
always be able to use intermediate buffers as we will have emptied them
before we need to communicate on that same dimension to process data for
another rank. Another possible schedule is to send close first, then
far, which could be easier to implement in some cases.

\begin{figure}
\centering
\includegraphics{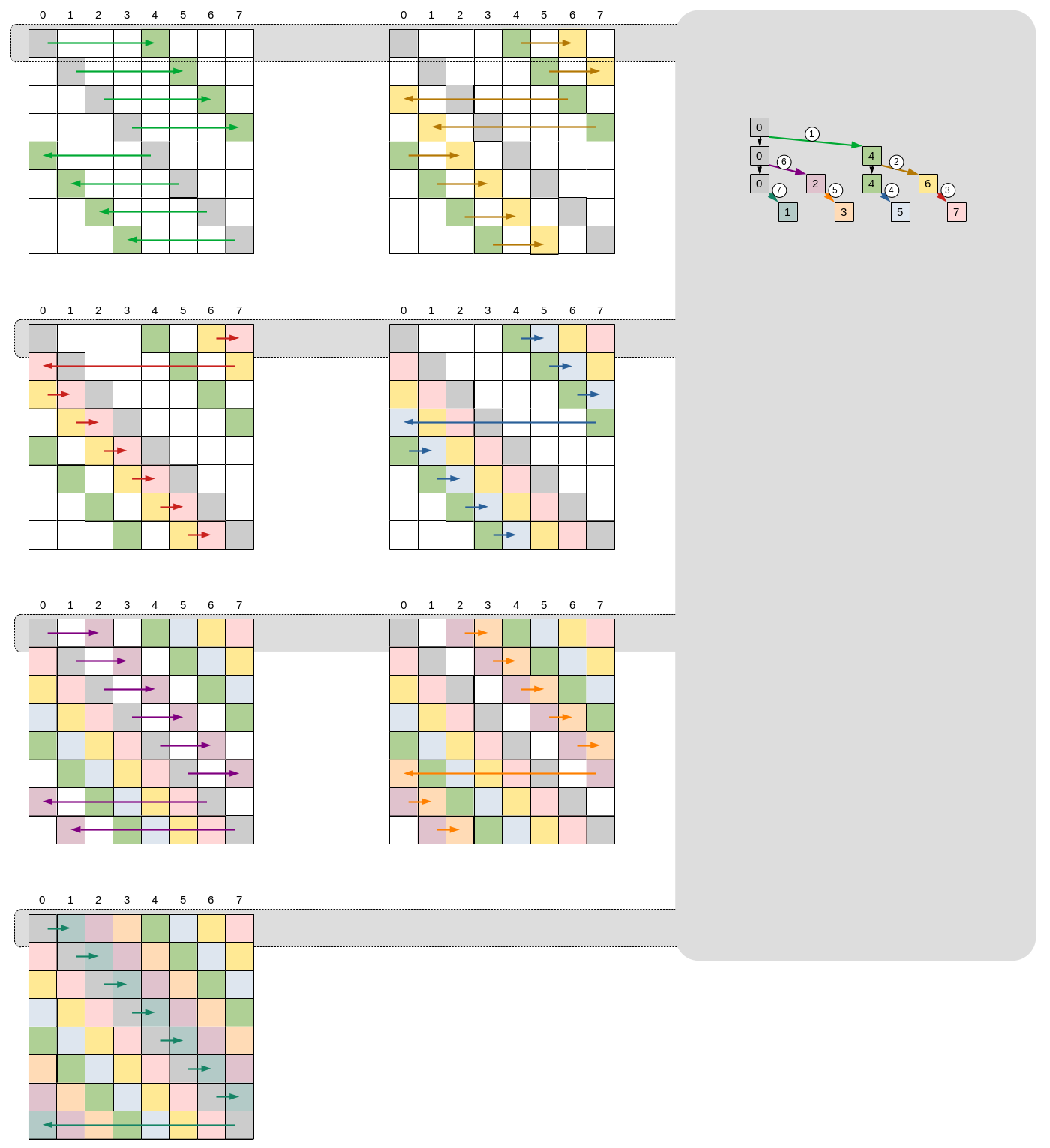}
\caption{PAT Tree with 8 ranks, fully linear}
\end{figure}

\hypertarget{conversion-to-reduce-scatter-operations}{%
\section{Conversion to reduce-scatter
operations}\label{conversion-to-reduce-scatter-operations}}

The algorithm was originally developed for reduce-scatter, as buffer
constraints are higher, with the idea that an algorithm which works for
reduce-scatter should also work for all-gather.

The reduce-scatter PAT algorithm works the same way as all-gather, but
with a reversed binomial tree.

\begin{figure}
\centering
\includegraphics{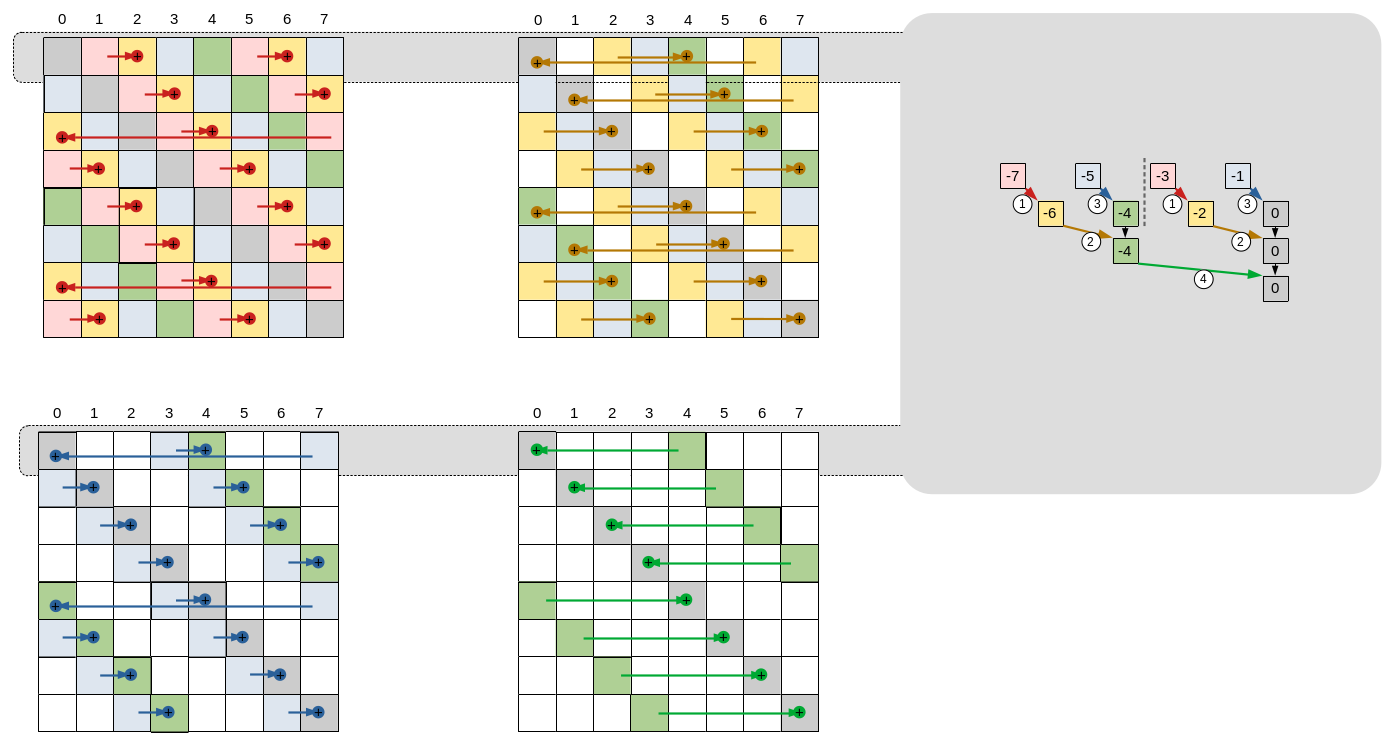}
\caption{PAT Tree for reduce-scatter}
\end{figure}

For reduce-scatter, we communicate with close dimensions first, then
far, and we also reverse the tree. Each time we receive data, we also
reduce it with the current accumulation buffer which will eventually be
sent to a higher dimension, except when we receive data for our own
buffer, in which case we accumulate in the user's receive buffer.

We also start with the parallel trees, and finish with the logarithmic
part of the tree (which is now the bottom).

\hypertarget{performance}{%
\section{Performance}\label{performance}}

The PAT algorithm implements a logarithmic number of steps in cases
where latency matters, and progressively increases the number of steps
as buffer size grows and constrains the amount we can store before
sending.

In nature, the algorithm is also linear on two fronts: the number of
chunks of data we need to manipulate separately is linear, given we
reversed the dimensions; and simply computing the steps is also a linear
operation. Hence, depending on the amount of optimization we can achieve
on those linear parts, and the network operation latency, the algorithm
may look linear or logarithmic. In any case, there is always a scale at
which the linear part will become predominant over the logarithmic part.
The performance factor over the ring algorithm will be dependent on how
much faster the linear part is, compared to the linear part of the ring.

The linear part of the PAT algorithm is purely local. It is therefore
CPU or GPU code, but given the complexity needed to implement the
algorithm, it may be significant.

The linear part of the \emph{Ring} algorithm is mostly related to
network operations, which are sent back-to-back, and therefore more
related to the message rate of the network than its latency.

\hypertarget{future-work}{%
\section{Future work}\label{future-work}}

The algorithm is implemented in NCCL 2.23 for 1 rank per node, as only
the inter-node part is implemented. It should be possible to implement
PAT algorithms with intra-node support however, as it is done in other
implementations, in particular in the collnet algorithms in NCCL. The
main challenge here is an additional level of complexity in already
complex algorithms.

As mentioned, other future work would also include further optimization
of the linear part. The complexity of the current algorithm is currently
high, and it is likely possible to simplify the algorithm to optimize it
further.

\hypertarget{references}{%
\section*{References}\label{references}}
\addcontentsline{toc}{section}{References}

\hypertarget{refs}{}
\leavevmode\hypertarget{ref-nccl}{}%
{[}1{]} ``NVIDIA collective communication library.''
\url{https://github.com/NVIDIA/nccl}.

\leavevmode\hypertarget{ref-bruck}{}%
{[}2{]} J. Bruck, C.-T. Ho, S. Kipnis, and D. Weathersby, ``Efficient
algorithms for all-to-all communications in multi-port message-passing
systems,'' in \emph{Proceedings of the sixth annual acm symposium on
parallel algorithms and architectures}, 1994, pp. 298--309.

\leavevmode\hypertarget{ref-thakur}{}%
{[}3{]} R. Thakur, R. Rabenseifner, and W. Gropp, ``Optimization of
collective communication operations in mpich,'' \emph{The International
Journal of High Performance Computing Applications}, vol. 19, no. 1, pp.
49--66, 2005.

\end{document}